\documentstyle[11pt,newpasp,twoside,epsf]{article}

\begin{document}

\title{The luminosity function of Palomar 5 and its tidal tails}
\author{A. Koch, M. Odenkirchen, E.K. Grebel, D. Mart\'{\i}nez-Delgado}
\affil{Max-Planck-Institut f\"ur Astronomie, K\"onigstuhl 17, D-69117 Heidelberg, Germany}
\author{J.A.R. Caldwell}
\affil{Space Telescope Science Institute, 3700 San Martin Drive, Baltimore, MD 21218, USA }

\begin{abstract}
We present the main sequence luminosity function of the
tidally disrupted globular cluster Palomar 5 and its tidal tails. 
For this work we analyzed imaging data obtained with the Wide Field
Camera at the INT (La Palma) and data from the Wide Field Imager at the MPG/ESO 2.2\,m 
telescope at La Silla down to a limiting magnitude of approximately 24.5\,mag in B.  
Our results indicate that preferentially fainter stars were removed from the cluster
so that the LF of the cluster's main body exhibits a
significant degree of flattening compared to other GCs.
This is attributed to its advanced dynamical evolution. 
The LF of the tails is, in turn, 
enhanced with faint, low-mass stars, which we interpret as a consequence
of mass segregation in the cluster.
\end{abstract}

\section{Introduction}
Palomar 5 is a faint ($M_V=-5.17$), sparse (M\,$\approx\,5000 M_{\sun}$) and low-concentration (c$\,=\,\log(r_t/r_c)\,=\,0.74$) 
globular cluster (GC) that is currently in a state of advanced dissolution due to the tidal forces
exerted by the Milky Way. Its dynamical evolution has been dominated by mass loss that
predominantly occured during passages through the Galactic disk. Direct evidence
for this disruption arises from the recent discovery of two tidal streams emanating from the
cluster's main body (Odenkirchen et al. 2001, 2003) which cover at least $\sim 10\deg$ on the sky.
In a study that used deep HST data Grillmair \& Smith (2001, hereinafter GS01) derived luminosity
functions (LFs) for two separate regions of the cluster's center and found them to be identical
to within their uncertainties. On the other hand they found Palomar 5 to be 
apparently depleted in low mass stars compared to other GCs. 

This work presents the first analysis of the stellar content of a GC's tidal debris, thus providing the possibility to 
directly test, whether a flattened LF can have its origin in tidal mass loss.
For this, we aim at measuring the luminosity function of the entire cluster and comparing it to that of the tidal tails.
Furthermore, if this deficiency is due to mass segregation through energy equipartition, one should expect the LF of the 
tidal streams to be enhanced in low-mass stars lost from the center.
\section{Data}
In order to distinguish the luminosity functions in different portions of the cluster we targeted
our deep imaging observations at the cluster's center and several regions in the tidal tail. 
We obtained data with two different wide field instruments, 
both of which provide a field of view of approximately $35\arcmin \times 35\arcmin$.
Firstly, we used the Wide Field Camera of the 2.5\,m INT at La Palma to obtain deep B and r' images.
These data reach a limiting magnitude of approximately 24.5\,mag in B.
Secondly, we observed Pal 5 using the Wide Field Imager at the 2.2\,m ESO/MPG telescope at La Silla, both in V and R. 
To estimate the contamination by field stars we also took images of two control regions located 
well away from the cluster (viz., 1$\fdg$5) and the tails, still representing a field typical for that area. The location of fields
is depicted in Fig. 1.
\begin{figure}
\plotfiddle{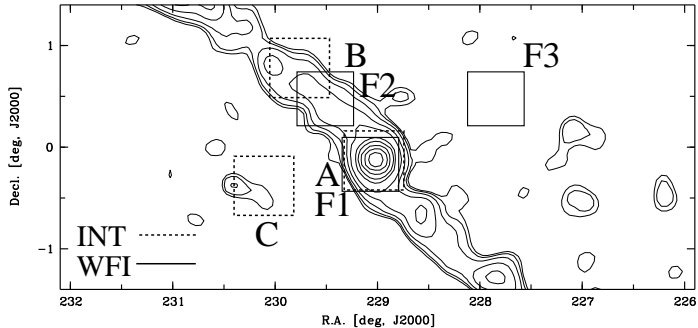}{3.8cm}{0}{35}{35}{-140}{0}
\caption{Observed fields overlaid on a contour plot of Palomar 5. Dashed squares mark the INT fields (labelled A-C), 
whereas the solid squares labelled F1-F3 denote our WFI observations.}
\end{figure}
\section{Color magnitude diagrams and number counts}
The resulting color magnitude diagrams are shown in Fig. 2.
The left panel shows stars within 3$\farcm$6 around the cluster center -- corresponding to the cluster's core radius.
Its  main sequence is well defined, whereas the horizontal and red giant branch are populated more sparsely.
There is also an apparent binary main sequence visible in the INT data, 
but these objects were not included in the analyses presented here.

A CMD of the tidal tail, on the other hand, is displayed in the right panel of Fig. 2. 
As the regions in the tails contain an increasing number of field stars, the main features of the cluster population 
within the tails are overlapped by the field population and do not stick out clearly .

In order to select primarily main sequence star candidates we calculated a 2$\sigma$-envelope around the main 
sequence as defined in Fig. 2, which was smoothed afterwards. 
\begin{figure}
\plottwo{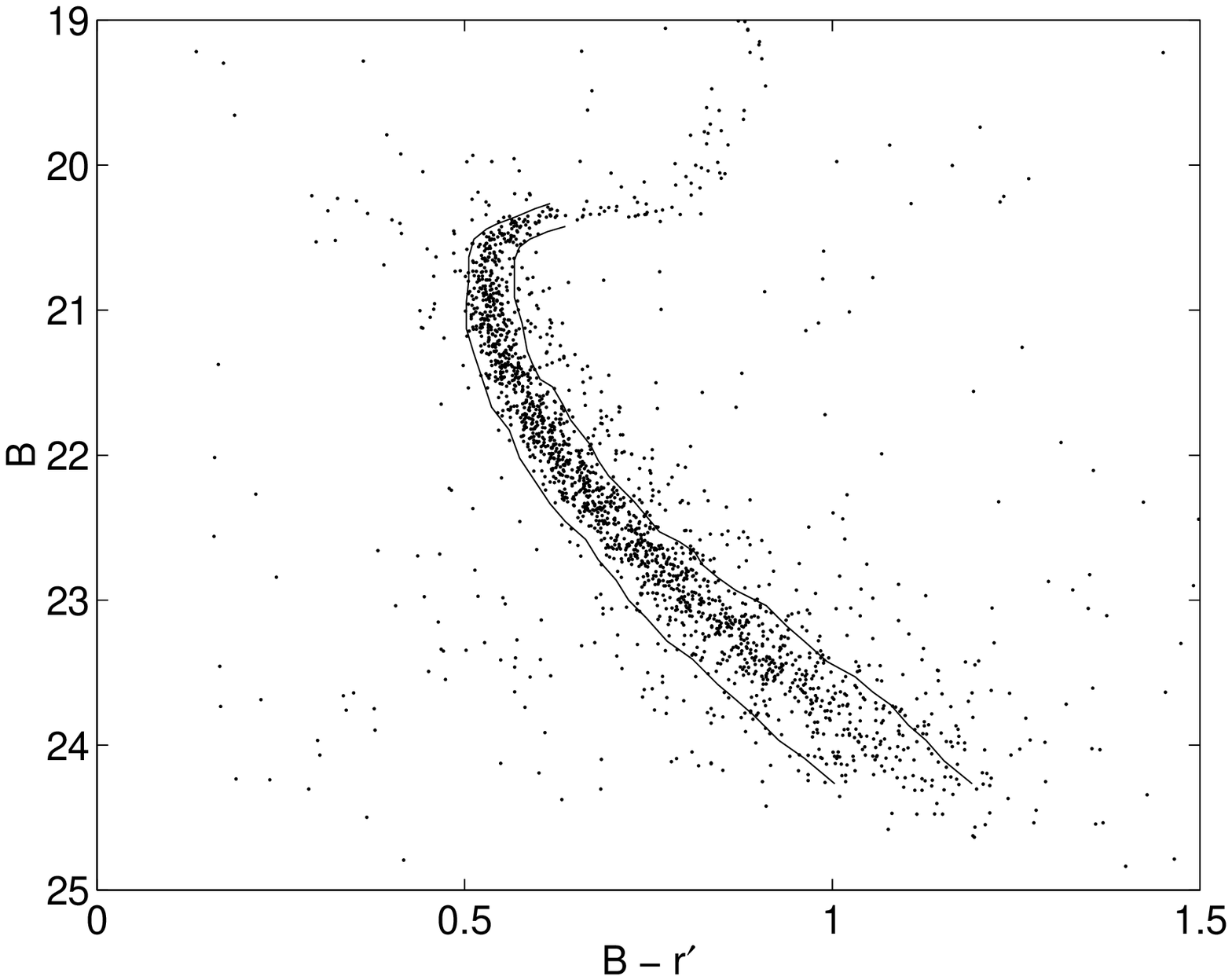}{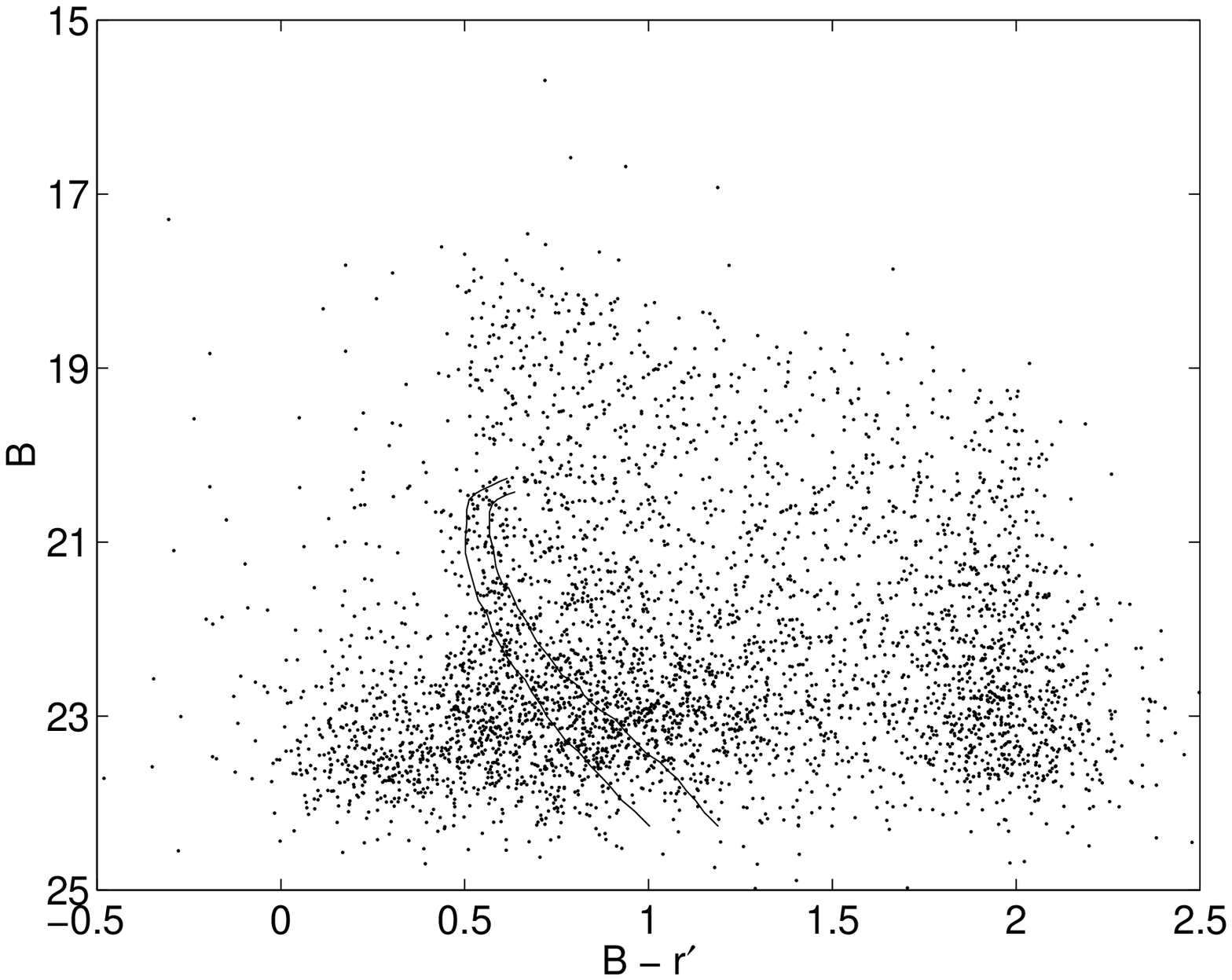}
\caption{Color magnitude diagram from INT data of Palomar 5's central 3$\farcm$6 ({\em left panel}), and the respective CMD for field B 
in the tidal tail ({\em right panel}).
The solid line is the 2$\sigma$-envelope used for number counts of main sequence stars.}
\end{figure}

Now for each observed field, using only objects within this 2$\sigma$-envelope, we performed number 
counts in bins of 0.5 mag. Field stars were statistically subtracted and
the resulting histogram was corrected for incompleteness effects
to construct the final stellar main sequence luminosity function ({\sc MSLF}). 

However, field corrections become increasingly uncertain the further one proceeds outwards, 
where our observed regions contain less cluster stars.
This makes it rather difficult to extract the LF of the tail population itself. 
\section{Cluster luminosity function} 
Fig.\,3 presents our {\sc MSLF}s of the cluster's central 3$\farcm$6 in comparison to the results for the northern tail. 
\begin{figure}
\plottwo{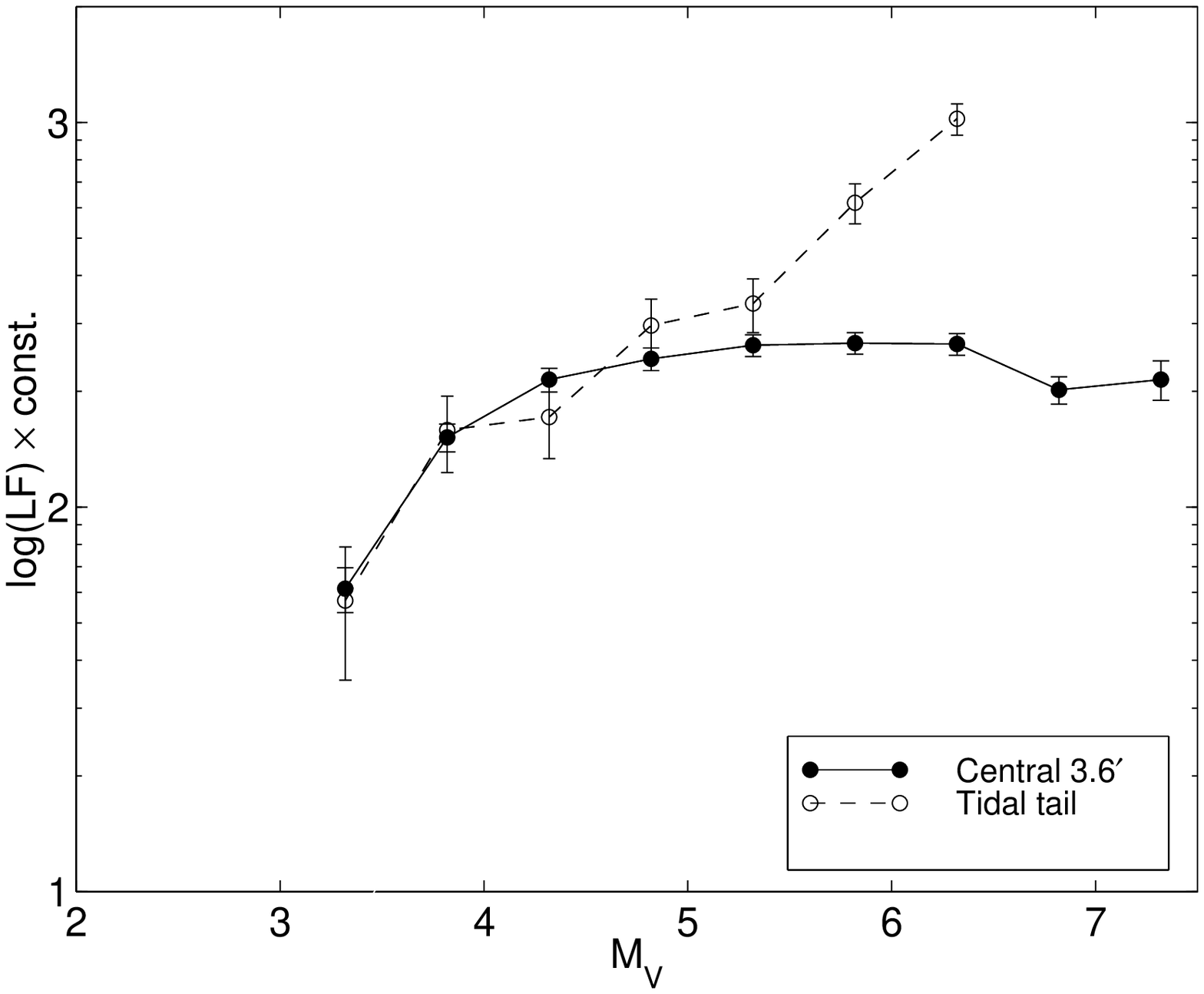}{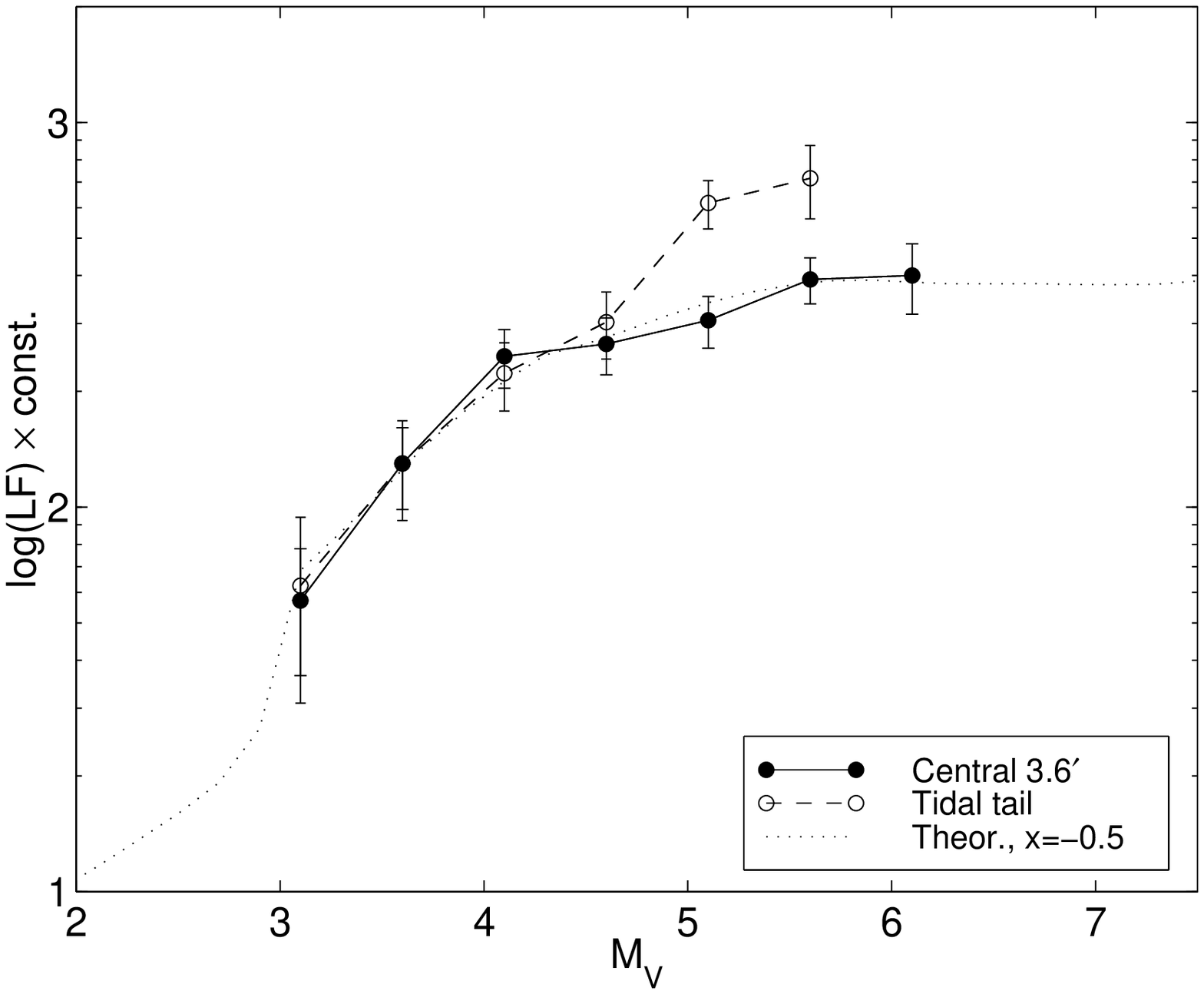}
\caption{Luminosity functions of Palomar 5's main body (solid line) compared to that from the tidal tail (dashed). INT data are shown left 
and the WFI results are displayed in the right panel.
The errorbars are purely statistical. The dotted line is a theoretical {\sc LF} based on a mass-luminosity-relation
 from BV92 using the remarkably low power-law exponent of $-$0.5. The curves were scaled to fit at the bright end.}
\end{figure}
The outstanding feature of these LFs is the high degree of flattening at the faint end 
which signifies deficiency in low mass stars.
This result is in good agreement with the HST study of a central portion of Pal 5 by GS01.
To quantify the flattening, we fit a theoretical LF to the data using
isochrones from Bergbusch \& VandenBerg (1992, BV92) with an age of 12\,Gyr and a metallicity of $-$1.48\,dex, both values close to Pal
5's parameters of 11.5\,Gyr (Martell, Smith, \& Grillmair 2002) and $-$1.43\,dex (Harris 1996).
The underlying mass function (MF) consists of a power-law with an exponent of $x\la -0.5$, which fit our data best.
This means that Pal 5's MF is remarkably shallow, because typical GCs are generally described by $x\la 0.3$ (Piotto \& Zoccali 1999). 
\section{Tidal tail luminosity function}
The {\sc MSLF} of the tidal tail, on the other hand, is consistent with the cluster's 
LF at the brighter end to within the uncertainties, but it
is apparently enhanced in low mass stars as can be deduced from a larger number of faint stars in 
the last magnitude bins. A quantitative measure of the enhancement can be given by the exponent from a power-law fit of 
the kind $N\propto M_V^{\alpha}$. The value of $\alpha$ was determined as
$0.6\,\pm\,0.2$ ($0.7\,\pm\,0.4$) for the central region and $4.0\,\pm\,0.3$ ($4.2\,\pm\,1.4$) for the tidal tail. 
These numbers refer to our INT data, whereas numbers in brackets were derived from the WFI data.
This  result confirms the depletion of the core of low mass stars and the complementary enhancement of the tails with
these stars. 
\section{Discussion}
Both our datasets from the INT and the WFI consistently show that the central portion of the 
low concentration cluster Palomar 5 is significantly deprived of low mass stars. These moved to the cluster's outskirts in the course
of dynamical mass segregation, where they were preferentially removed through evaporation and increasingly by tidal stripping.
Presently, the faint stars are predominantly found in the tidal tails, which confirms the segregation scenario.
However, it cannot be entirely ruled out that the distinctness of this effect is biassed to a certain extent by an increasing field contamination.

Since MS is generally believed to be negligible in low concentration GCs (Pryor, Smith, \& McClure 1986), 
it seems that Pal 5 must have undergone severe strucural changes during its evolution.

\end{document}